\documentclass[%
 aip,
 amsmath,amssymb,
 reprint,%
]{revtex4-1}

\usepackage{graphicx}
\usepackage{dcolumn}
\usepackage{bm}
\usepackage{url}
\usepackage{hyperref}
\usepackage[utf8]{inputenc}
\usepackage[T1]{fontenc}
\usepackage{mathptmx}
\usepackage{etoolbox}\usepackage{comment}
\usepackage{bm}
\usepackage[dvipsnames]{xcolor}
\usepackage{graphicx}
\usepackage{dcolumn}
\usepackage{bm}

\makeatletter
\def\@email#1#2{%
 \endgroup
 \patchcmd{\titleblock@produce}
  {\frontmatter@RRAPformat}
  {\frontmatter@RRAPformat{\produce@RRAP{*#1\href{mailto:#2}{#2}}}\frontmatter@RRAPformat}
  {}{}
}%
\makeatother
\begin{document}

\preprint{AIP/123-QED}


\title{Shaping terahertz harmonic frequency combs with frequency dependent external reflectors}
\author{Carlo Silvestri}
\affiliation{Institute of Photonics and Optical Science (IPOS), School of Physics, University of Sydney, Sydney, New South Wales 2006, Australia
}
\affiliation{School of Electrical Engineering and Computer Science, The University of Queensland, Brisbane, QLD 4072, Australia
}

\author{Xiaoqiong Qi}%
\affiliation{School of Electrical Engineering and Computer Science, The University of Queensland, Brisbane, QLD 4072, Australia
}
\author{Thomas Taimre}
\affiliation{School of Mathematics and Physics, The University of Queensland, Brisbane, QLD 4072, Australia
}

\author{Aleksandar D. Raki\'c}
\affiliation{School of Electrical Engineering and Computer Science, The University of Queensland, Brisbane, QLD 4072, Australia
}

\date{\today}

\begin{abstract}
We present a method for engineering harmonic frequency combs (HFCs) in the terahertz spectral region. This approach involves interfacing a quantum cascade laser (QCL) with an external reflector featuring frequency-dependent reflectivity. A notable advantage of this method over existing ones is its dual functionality in shaping HFCs, allowing for control over both the frequency offset and comb spacing based on the external reflectivity profile. Moreover, the resulting HFCs manifest as sequences of short pulses in the time domain.  Consequently, our method enables the generation of ultrashort picosecond pulses passively, providing a distinct alternative to conventional pulse generation systems reliant on active bias current modulation, which struggle with modulation frequencies significantly higher than the first beatnote. This offers intriguing prospects for utilizing HFCs in pump and probe spectroscopy, a field already recognized in the literature as one of the most compelling applications of these states.  Furthermore, we demonstrate that these HFCs can be triggered from an initial condition of free-running unlocked dynamics, eliminating the need for assuming free-running comb emission. Thus, the utilization of an external, frequency-dependent reflector is capable of enhancing the coherence of the QCL emission. 
\end{abstract}

\maketitle

\begin{figure}[t]
   \begin{center}
   \includegraphics[width=0.5\textwidth]{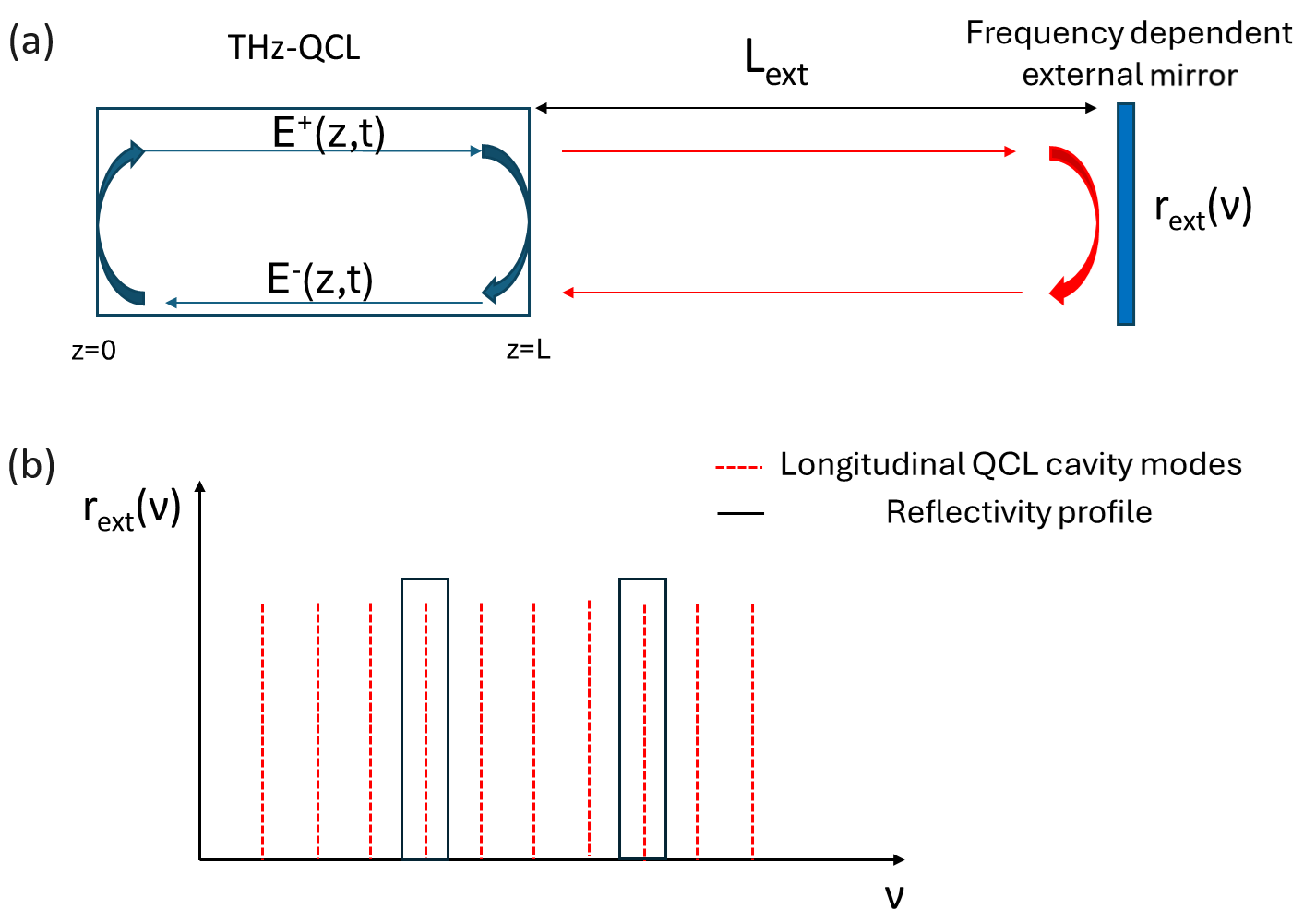}
   \end{center}
   \caption{(a) Self-mixing configuration with a FP THz QCL and a frequency-dependent mirror. (b) Reflectivity profile of the mirror in the frequency domain (black lines), characterized by two frequency pulses, each aligned with a longitudinal cavity mode of the QCL (red dashed lines).} \label{fig0}
\end{figure}

The discovery of the spontaneous formation of optical frequency combs (OFCs) in quantum cascade lasers  (QCLs) \cite{Hugi2012,Burghoff2014} has brought significant benefits for applications in the fields of spectroscopy and optical communication in the mid-infrared (mid-IR) and terahertz (THz) spectral regions. \cite{SilvestriReview,PiccardoReview,Faist_2016,Villares_2014,Corrias} More recently, it has been demonstrated that not only fundamental frequency combs are emitted by QCLs, but also harmonic frequency combs (HFCs), characterized by an optical line spacing equal to a multiple of the free-spectral range (FSR) of the laser cavity.\cite{Kazakov21,ForrerHFC,Dhillon1,Kazakov2017} The importance of HFCs stems from their potential for a wide range of applications, encompassing microwave generation, pump and probe spectroscopy, and wireless communication.\cite{PiccardoHFCOptex} Consequently, considerable efforts have been directed towards developing methods to engineer and control these states. Studies have demonstrated the feasibility of creating and controlling HFCs through the utilization of electrical \cite{Dhillon1,Silvestri23PRA} and optical injection.\cite{PiccardoOptical} techniques\\
Recently, methods to engineer HFCs based on placing defects on the top surface of a THz QCL have been proposed.\cite{VitielloHFC1,VitielloHFC2} Furthermore, we presented an alternative method for inducing and manipulating HFCs through interferometric optical feedback. This approach involves interfacing the QCL with a frequency-independent external reflector in a self-mixing configuration, offering a significant simplification of the experimental setup compared to other techniques.\cite{Silvestrifb} Our investigation led to the theoretical prediction of tuning the harmonic order of a frequency comb, and thereby the spacing between the optical lines, especially in a regime with a short external cavity (on the order of mm). In this scenario, the introduction of an external reflector creates a three-mirror cavity with a modified spectral behavior compared to the two-mirror laser cavity without feedback. However, as previously mentioned, wide tunability in this setup is only guaranteed within the short external cavity limit, where the free-spectral range (FSR) of the external cavity is of the same order of magnitude as that of the laser cavity. Additionally, while this method allows for tuning the comb spacing, it does not impact the offset frequency of the comb.\\
In this work, we present a novel study on the interaction between QCL combs and external optical feedback, this time using a reflector with frequency-dependent reflectivity. We demonstrate how, in this new scenario, it is possible to shape HFCs without the constraint of the short cavity limit. In this case, it is not the effect of the external cavity length that triggers HFCs, but rather the presence of a mirror with high reflectivity at specific frequencies. This allows the support of only some of the QCL's longitudinal modes, leading the selection of the comb spacing and triggering the emission of harmonic states. Additionally, this method allows to select both the spacing and offset of the comb, based on the frequency dependence of the external reflectivity. The studied configuration is presented in Fig.~\ref{fig0}.

A FP THz-QCL capable of emitting OFCs is interfaced with a mirror with frequency-dependent reflectivity $r_\mathrm{ext}(\nu)$ in a self-mixing scheme, with a distance $L_\mathrm{ext}$ between the exit facet of the QCL and the external reflector (Fig.~\ref{fig0}(a)). $E^+$ and $E^-$ represent the forward and backward fields circulating in the FP cavity, respectively. The reflectivity profile in the frequency domain is illustrated in Fig.~\ref{fig0}(b) (indicated by the black lines), consisting of two regions with non-zero reflectivity corresponding to frequency pulses, each aligned with one of the longitudinal modes of the QCL cavity. A similar reflectivity profile could be achieved by utilizing optical components such as THz frequency band-pass filters \cite{Paul09} placed in front of a mirror frequency independent in the THz range.
\\\\
The laser dynamics in the setup depicted in Fig.~\ref{fig0} are simulated using a full set of effective semiconductor Maxwell--Bloch equations (ESMBEs) \cite{Columbo2018,columbo2007,Silvestri22,SilvestriThesis} for a Fabry--Perot (FP) THz QCL. The device parameters employed for the simulations are provided in the supplementary materials. Optical feedback is incorporated by adding the term $E_\mathrm{FB}(t)$ to the free-running boundary conditions of the ESMBEs:
\begin{eqnarray}
E^-(L, t)&=&\sqrt{R}E^+(L,t)+E_\mathrm{FB}(t),\label{bc2fb1}\\
E^+(0, t)&=&\sqrt{R}E^-(0,t)\label{bc2fb2}
\end{eqnarray}\\
where $L$ represents the laser cavity length and $R$ denotes the facet reflectivity of the QCL. In presence of a frequency-dependent external mirror with reflectivity $r_{\mathrm{ext}}(\nu)$, the total reflection coefficient associated with the considered configuration is expressed as:
\begin{eqnarray}
r(\nu)=r_\mathrm{ext}\left(\nu\right)t_\mathrm{L}^2e^{-i\Phi_\mathrm{ext}}e^{-i2\pi(\nu-\nu_0)\tau_\mathrm{ext}}\label{r_omega}
\end{eqnarray}
where $t_\mathrm{L}$ represents the facet transmissivity, $\Phi_\mathrm{ext}=4\pi L_\mathrm{ext}\nu_0/c$, $\tau_\mathrm{ext}=2L_\mathrm{ext}/c$, $L_\mathrm{ext}$ is the external cavity length, and $\nu_0$ is the central emission frequency. In this scenario, the feedback field $E_\mathrm{FB}(t)$ is determined by multiplying the forward field $E^+(L,\nu)$ by $r(\nu)$ as given by Eq.~(\ref{r_omega}), and then inverse-transforming to return to the time domain.

\begin{figure*}[t]
   \begin{center}
   \includegraphics[width=1\textwidth]{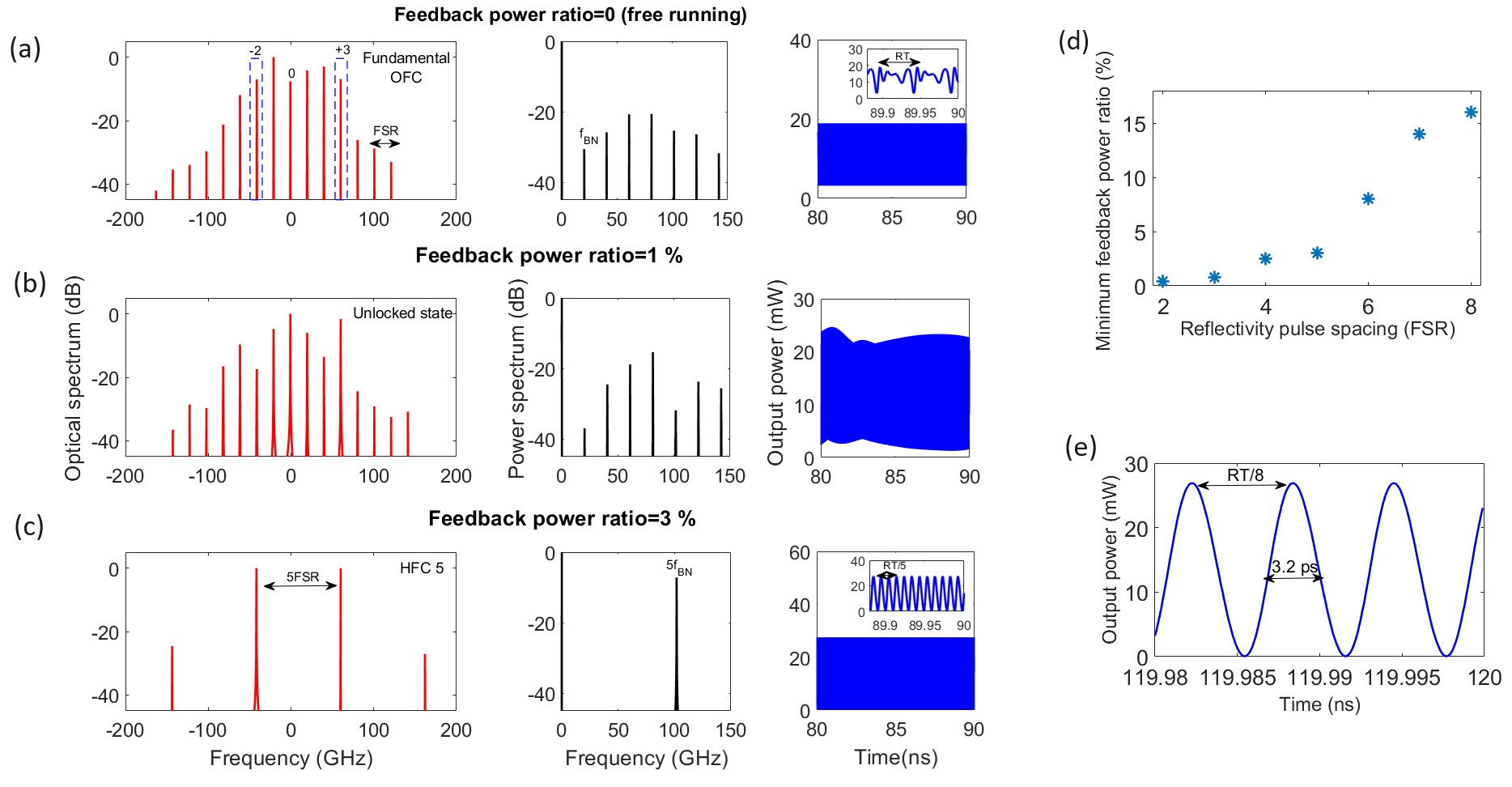}
   \end{center}
   \caption{(a) Optical spectrum (left), power spectrum (center), and power versus time trace of the free-running dense OFC used as the initial condition. The blue dashed lines superimposed on the optical spectrum outline the reflectivity profile of the reflector in the frequency domain, characterized by two pulses corresponding to the OFC modes $-2$ and +3; the inset in the power temporal trace shows the regular repetition of a multipeak structure each roundtrip time (RT). The same figures of merit are shown for the irregular regime obtained for feedback power ratio $\epsilon^2=1~\%$ (b), and for the 5-th order HFC reported for $\epsilon^2=3~\%$ (c). (d) Minimum feedback power ratio for HFC formation as a function of reflector pulse spacing. (e) Power trace for the 8-th order HFC of Fig.~(d), exhibiting a pulse width of 3.2~ps. } \label{fig1}
\end{figure*}
In the absence of feedback, we consider the initial condition to be the free-running emission of a dense OFC by the QCL. This state, illustrated in Fig.~\ref{fig1}(a), naturally exhibits equispaced modes with spacing corresponding to the QCL FSR$=20.8~$GHz (left panel), a narrow peak at the first beatnote frequency $f_\mathrm{BN}$ (centre), and a regular temporal evolution of the output power (right). The free-running OFC emission is achieved by setting the laser bias current $I = 1.5I_{\text{thr}}$, where $I_{\text{thr}}$ represents the QCL threshold current.

The introduction of optical feedback implies a change in the laser emission. In all the numerical experiments presented in this study, feedback is implemented by placing a mirror at a distance of $L_\mathrm{ext}=10~$cm from the QCL's exit facet. This external mirror is characterized by a frequency-dependent reflectivity profile, featuring two distinct frequency-domain pulses. Each pulse spans 10~GHz, roughly half the FSR, and its central frequency aligns with specific modes of the original comb. In the result illustrated in Figs.~\ref{fig1}(a)--(c), these pulses correspond to modes $-2$ and +3 of the original comb, with mode 0 referring to the central comb mode. In Fig.~\ref{fig1}(a), the dashed blue lines precisely outline the reflectivity profile in the frequency domain. Therefore, if $\epsilon$ represents the reflectivity peak value of each pulse, and $\nu_\mathrm{n_1}$, $\nu_\mathrm{n_2}$ are the central frequencies of the two frequency pulses, corresponding respectively to the QCL longitudinal modes $n_1$ and $n_2$, $r(\nu)$ can be expressed as:\\
\begin{eqnarray}
                r_\mathrm{ext}\left(\nu\right) =  
        \begin{cases}
        \epsilon &\quad\text{if $\nu_\mathrm{n_1}-5~\mathrm{GHz}$}\le \text{$\nu$}\le \text{$\nu_\mathrm{n_1}+5~\mathrm{GHz}$}\\
        \epsilon &\quad\text{if $\nu_\mathrm{n_2}-5~\mathrm{GHz}$}\le \text{$\nu$}\le \text{$\nu_\mathrm{n_2}+5~\mathrm{GHz}$}\\
        0 &\quad\text{elsewhere.} \\
            \end{cases}\nonumber\\
\end{eqnarray}
We assume the same feedback strength, denoted as $\epsilon$, for both frequency pulses. The term $\epsilon^2$ represents the corresponding spectrally local feedback power ratio.\\
In the scenario depicted in Figs.~\ref{fig1}(a)--(c), the central frequencies of the two pulses are separated by 5~FSR. When $\epsilon^2 = 1\%$, a transition is observed from the free-running OFC of Fig. \ref{fig1}(a) to an unlocked multimode dense state (Fig.~\ref{fig1}(b)), as evident from both the optical and power spectra, and from the analysis of the output power plot (right panel), revealing irregular amplitude modulations indicative of unlocked dynamics. By further increasing the feedback power to 3\%, a 5th-order HFC is induced, with the highest peaks corresponding to modes $-2$ and +3 of the original comb, i.e., those supported by the external reflector. Therefore, for an experimentally achievable feedback ratio (3~\%), the spectral features of the reflector are impressed upon the comb dynamics, resulting in a harmonic comb state with spacing determined by the mirror reflectivity. This phenomenon differs from what was presented in ref.~\cite{Silvestrifb}. In that case, the spectral effective reflectivity of the right QCL facet in the frequency domain was modified by tuning the external cavity length. Here, while keeping the external cavity length fixed, we introduce spectrally dependent reflectivity, leading to the reinjection of only certain original comb modes (in this instance, $-2$ and +3). Consequently, above a certain feedback level, this reinjection promotes these modes at the expense of the intermediate optical lines. In the presence of the strong nonlinearities typical of QCLs (included in the ESMBEs model), the promoted modes are able to lock, producing a harmonic comb.\\
We replicate this experiment by setting different values of the spacing between the frequency pulses, corresponding to multiples of the cavity FSR. We follow this criterion:
\begin{itemize}
\item For a spacing corresponding to $n$ FSR with $n$ even: we place the pulses in correspondence with modes $-\frac{n}{2}$ and $+\frac{n}{2}$. For example, if $n=4$, we consider modes $-2$ and $+2$.
\item For $n$~FSR with $n$ odd (therefore $n=k+1$ with $k$ even): we place the notches at modes $-\frac{k}{2}$ and $+\frac{k}{2}+1$. For example, if $n=3$, we consider modes $-1$ and $+2$.
\end{itemize}
\begin{figure}[t]
   \begin{center}
   \includegraphics[width=0.43\textwidth]{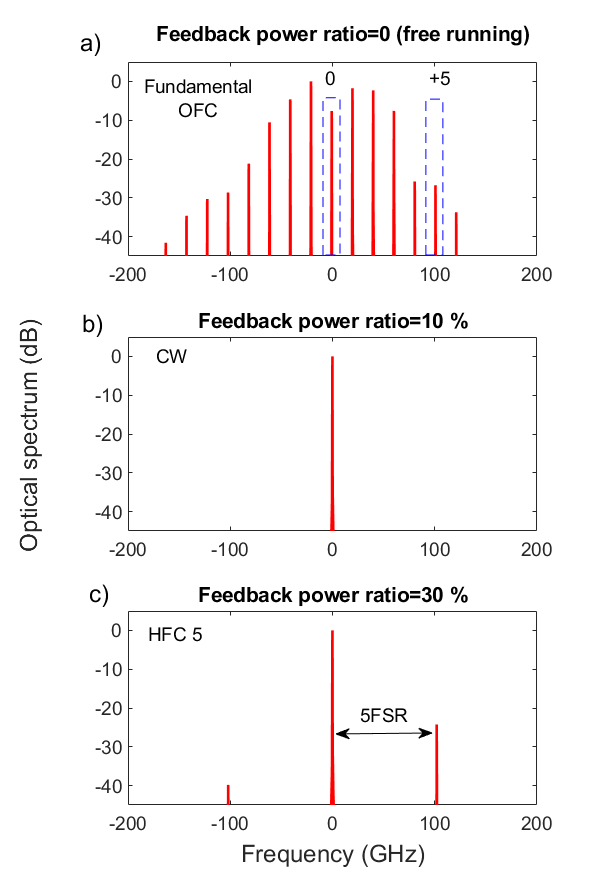}
   \end{center}
   \caption{Optical spectrum for the free-running OFC state (a), the continuous wave (CW) for $\epsilon^2=10~\%$ (b), and the 5th-order HFC for $\epsilon^2=30~\%$ (c). In Fig.~(a), the reflectivity profile is shown (dashed lines), with frequency pulses aligned with modes 0 and +5 of the free-running OFC.} \label{fig2}
\end{figure}
This arrangement was chosen to ensure a symmetric placement of the reflectivity peaks relative to the center of the gain curve (mode 0). This assumption will be modified later, where we will discuss how feedback strength required to achieve the HFC states changes in the presence of a misalignment relative to the gain curve. We consider $n=2,\ldots,8$, measuring the minimum feedback power ratio needed to induce an $n$-th order HFC. The results are plotted in Fig.~\ref{fig1}(d). We notice that the minimum feedback level increases with the reflectivity pulse spacing. This indicates the importance of the QCL gain curve in the formation of these HFCs. In fact, as we increase the displacement of the pulses from the central frequency, where the gain is maximum (as assumed in the ESMBEs model used \cite{Columbo2018,Silvestri20}), greater feedback is required to induce the harmonic combs. This is because, as the distance from the center increases, the modes experience progressively lower gain. For $n \geq 9$, no $n$-th order HFC can be induced within $\epsilon^2<30$ \%. For this reason, we considered $n\leq 8$ in Fig.~\ref{fig1}(d).\\
Figure \ref{fig1}(e) displays the HFC with highest order observed in this part of the study, an 8th order state. Examining the temporal power profile for this HFC reveals a sequence of pulses each 3.2 ps wide. Thus, with a passive method, i.e. without radiofrequency injection, we induce short pulses on the picosecond scale. The potential to generate ultra-short pulses associated with high-order HFCs was outlined in ref.\cite{PiccardoHFCOptex}, emphasizing potential applications for pump and probe spectroscopy. Our proposed method for shaping HFCs thus enhances prospects for such applications. Furthermore, we verified that if the frequency pulses are placed between two adjacent longitudinal modes, the QCL emission remains unmodified by the presence of the reflector.

\begin{figure}[t]
   \begin{center}
   \includegraphics[width=0.45\textwidth]{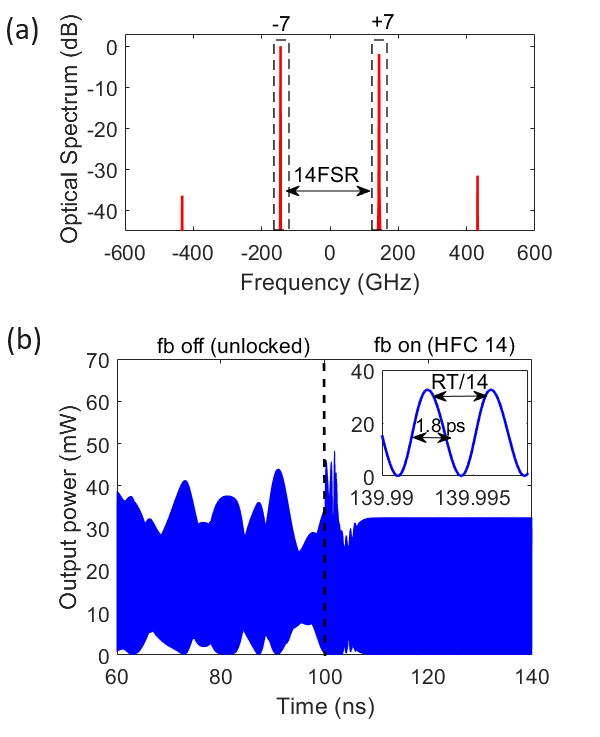}
   \end{center}
   \caption{Formation of a high-order HFC in a QCL with a gain bandwidth of 1.6 THz. (a) Optical spectrum of a 14th-order HFC generated by utilizing an external reflector with frequency pulses corresponding to the QCL longitudinal modes $-7$ and +7; the reflectivity spectral profile is represented with dashed lines. (b) Temporal evolution of the output power without feedback (to the left of the dashed vertical line), where the dynamics is irregular, and with feedback (to the right of the dashed line), where coherent dynamics is observed. The inset shows a zoomed-in view of the temporal trace, revealing consecutive pulses with a repetition period of RT/14 and a full width at half maximum of $1.8~$ps.} \label{fig3}
\end{figure}

To delve deeper into the influence of the reflectivity peak positions relative to the gain peak, we replicate the numerical experiment depicted in Figs.~\ref{fig1}(a)--(c), this time positioning the frequency pulses between mode 0 (central) and mode +5 of the free-running comb, rather than between mode $-2$ and mode +3. The initial condition is the same free-running OFC considered in Fig.~\ref{fig1}(a), with its optical spectrum again presented in Fig.~\ref{fig2}(a). We observe a transition to a single mode for intermediate feedback ratios between 0 and 30 \%, as depicted in Fig.~\ref{fig2}(b) for $\epsilon^2=10$ \%. Subsequently, a 5th-order HFC forms for $\epsilon=30$ \%. Comparing with Fig.~\ref{fig1}(c), we note that the transition to HFC 5 occurred at $\epsilon^2$=3 \%, which is one order of magnitude lower than the case of Fig.~\ref{fig2}(c). This difference arises from the spacing of the modes reinjected by the reflector. In the former case, these modes are spaced by 2~FSR and 3~FSR from the central one, whereas in the latter case, we consider a re-injected mode spaced by 5~FSR from mode 0. Notably, mode +5 experiences low gain (as evidenced by its 30 dB suppression ratio with respect to the highest peak in the free-running OFC of Fig.~\ref{fig2}(a)), necessitating higher reinjection to make it dominant over the intermediate ones, thereby enabling the observation of a HFC. From this comparison, two insights emerge. Firstly, besides the spacing between the reflectivity pulses in the frequency domain, their position relative to the center of the free-running comb is also crucial. When the non-zero reflectivity regions are symmetrically or nearly symmetrically arranged relative to the center of the comb, it becomes easier to obtain a harmonic comb with the same spacing. Secondly, by comparing HFC 5 of Fig.~\ref{fig1}(c) with that of Fig.~\ref{fig2}(c), we notice a difference in frequency offset despite having the same spacing. This indicates that the proposed method for shaping harmonic combs not only determines the spacing but also controls the frequency offset, offering complete customization of the generated combs.\\
At this point, we examine whether with a gain bandwidth exceeding 1 THz, characteristic of QCLs with multi-stack active regions, it is possible to obtain higher-order harmonic combs than those shown so far. We set a value of 1.6 THz, keeping  $I=1.5I_\mathrm{thr}$ as utilized in Fig.~\ref{fig1}, and using a reflector with pulses corresponding to the longitudinal modes $-7$ and +7. With a feedback power ratio of 45 \%, we are able to induce a 14th order HFC, as illustrated in Fig.~\ref{fig3}(a). However, an interesting outcome of this numerical experiment lies in the time domain scenario. Indeed, as observed in Fig.~\ref{fig3}(b), during free-running operation, the laser emits an irregular unlocked regime, and when the feedback is activated, it transitions to a HFC regime. This indicates that the feedback is capable of inducing a transition from an unlocked to a locked state. Evidences on the possibility of improving the coherence level of a QCL multimode state and the stability of a OFC in presence of an external mirror were experimentally reported.\cite{comb_stability,Hillbrand18} Furthermore, it can be observed in the inset of Fig.~\ref{fig3}(b) how the 14-th order HFC is associated with ultrashort pulses of 1.8 ps duration. This demonstrates that the use of multi-stack QCLs combined with a frequency dependent external reflector can lead to an improvement of the proposed configuration in terms of ultrashort pulse generation. This would potentially replace active mode locking based on radiofrequency injection, which would not be achievable at such high modulation frequencies.\cite{Dhillon1}

In conclusion, we have explored the interaction between QCL combs and external optical feedback provided by a frequency-dependent mirror. This investigated configuration facilitates the generation of harmonic combs, leveraging the characteristics of the external reflector in the frequency domain. This allows for precise control over both the offset and spacing of the combs. We have observed how the position of the reflectivity peaks of the mirror, responsible for the re-injection of only certain optical lines of the free-running emission, relative to the center of the QCL gain curve, plays a crucial role in determining the minimum feedback required to trigger the HFCs. Furthermore, as higher-order harmonic combs correspond to sequences of pulses in the time domain, the proposed method also serves as a system for passively generating ultrashort pulses. It is noteworthy that for typical gain curve values of QCLs with multi-stack active regions, higher than those of conventional QCLs, it is possible to induce higher-order HFCs, corresponding to pulses with widths below 2 ps.
\section*{Acknowledgment}
The authors acknowledge the funding from the Australian Research Council Discovery Project (Grant No. DP200101948).
\section*{Author Declarations}
The authors have no conflicts to disclose.
\section*{Data Availability}
The data that support the findings of this study are available from the corresponding author upon reasonable request. 
\section*{References}
\bibliography{aipsamp}

\begin{thebibliography}{25}%
\makeatletter
\providecommand \@ifxundefined [1]{%
 \@ifx{#1\undefined}
}%
\providecommand \@ifnum [1]{%
 \ifnum #1\expandafter \@firstoftwo
 \else \expandafter \@secondoftwo
 \fi
}%
\providecommand \@ifx [1]{%
 \ifx #1\expandafter \@firstoftwo
 \else \expandafter \@secondoftwo
 \fi
}%
\providecommand \natexlab [1]{#1}%
\providecommand \enquote  [1]{``#1''}%
\providecommand \bibnamefont  [1]{#1}%
\providecommand \bibfnamefont [1]{#1}%
\providecommand \citenamefont [1]{#1}%
\providecommand \href@noop [0]{\@secondoftwo}%
\providecommand \href [0]{\begingroup \@sanitize@url \@href}%
\providecommand \@href[1]{\@@startlink{#1}\@@href}%
\providecommand \@@href[1]{\endgroup#1\@@endlink}%
\providecommand \@sanitize@url [0]{\catcode `\\12\catcode `\$12\catcode `\&12\catcode `\#12\catcode `\^12\catcode `\_12\catcode `\%12\relax}%
\providecommand \@@startlink[1]{}%
\providecommand \@@endlink[0]{}%
\providecommand \url  [0]{\begingroup\@sanitize@url \@url }%
\providecommand \@url [1]{\endgroup\@href {#1}{\urlprefix }}%
\providecommand \urlprefix  [0]{URL }%
\providecommand \Eprint [0]{\href }%
\providecommand \doibase [0]{http://dx.doi.org/}%
\providecommand \selectlanguage [0]{\@gobble}%
\providecommand \bibinfo  [0]{\@secondoftwo}%
\providecommand \bibfield  [0]{\@secondoftwo}%
\providecommand \translation [1]{[#1]}%
\providecommand \BibitemOpen [0]{}%
\providecommand \bibitemStop [0]{}%
\providecommand \bibitemNoStop [0]{.\EOS\space}%
\providecommand \EOS [0]{\spacefactor3000\relax}%
\providecommand \BibitemShut  [1]{\csname bibitem#1\endcsname}%
\let\auto@bib@innerbib\@empty
\bibitem [{\citenamefont {Hugi}\ \emph {et~al.}(2012)\citenamefont {Hugi}, \citenamefont {Villares}, \citenamefont {Blaser}, \citenamefont {Liu},\ and\ \citenamefont {Faist}}]{Hugi2012}%
  \BibitemOpen
  \bibfield  {author} {\bibinfo {author} {\bibfnamefont {A.}~\bibnamefont {Hugi}}, \bibinfo {author} {\bibfnamefont {G.}~\bibnamefont {Villares}}, \bibinfo {author} {\bibfnamefont {S.}~\bibnamefont {Blaser}}, \bibinfo {author} {\bibfnamefont {H.~C.}\ \bibnamefont {Liu}}, \ and\ \bibinfo {author} {\bibfnamefont {J.}~\bibnamefont {Faist}},\ }\bibfield  {title} {\enquote {\bibinfo {title} {Mid-infrared frequency comb based on a quantum cascade laser},}\ }\href {\doibase 10.1038/nature11620} {\bibfield  {journal} {\bibinfo  {journal} {Nature}\ }\textbf {\bibinfo {volume} {492}},\ \bibinfo {pages} {229--233} (\bibinfo {year} {2012})}\BibitemShut {NoStop}%
\bibitem [{\citenamefont {Burghoff}\ \emph {et~al.}(2014)\citenamefont {Burghoff}, \citenamefont {Kao}, \citenamefont {Han}, \citenamefont {Chan}, \citenamefont {Cai}, \citenamefont {Yang}, \citenamefont {Hayton}, \citenamefont {Gao}, \citenamefont {Reno},\ and\ \citenamefont {Hu}}]{Burghoff2014}%
  \BibitemOpen
  \bibfield  {author} {\bibinfo {author} {\bibfnamefont {D.}~\bibnamefont {Burghoff}}, \bibinfo {author} {\bibfnamefont {T.-Y.}\ \bibnamefont {Kao}}, \bibinfo {author} {\bibfnamefont {N.}~\bibnamefont {Han}}, \bibinfo {author} {\bibfnamefont {C.~W.~I.}\ \bibnamefont {Chan}}, \bibinfo {author} {\bibfnamefont {X.}~\bibnamefont {Cai}}, \bibinfo {author} {\bibfnamefont {Y.}~\bibnamefont {Yang}}, \bibinfo {author} {\bibfnamefont {D.~J.}\ \bibnamefont {Hayton}}, \bibinfo {author} {\bibfnamefont {J.-R.}\ \bibnamefont {Gao}}, \bibinfo {author} {\bibfnamefont {J.~L.}\ \bibnamefont {Reno}}, \ and\ \bibinfo {author} {\bibfnamefont {Q.}~\bibnamefont {Hu}},\ }\bibfield  {title} {\enquote {\bibinfo {title} {Terahertz laser frequency combs},}\ }\href {\doibase 10.1038/nphoton.2014.85} {\bibfield  {journal} {\bibinfo  {journal} {Nature Photonics}\ }\textbf {\bibinfo {volume} {8}},\ \bibinfo {pages} {462--467} (\bibinfo {year} {2014})}\BibitemShut {NoStop}%
\bibitem [{\citenamefont {Silvestri}\ \emph {et~al.}(2023{\natexlab{a}})\citenamefont {Silvestri}, \citenamefont {Qi}, \citenamefont {Taimre}, \citenamefont {Bertling},\ and\ \citenamefont {Rakić}}]{SilvestriReview}%
  \BibitemOpen
  \bibfield  {author} {\bibinfo {author} {\bibfnamefont {C.}~\bibnamefont {Silvestri}}, \bibinfo {author} {\bibfnamefont {X.}~\bibnamefont {Qi}}, \bibinfo {author} {\bibfnamefont {T.}~\bibnamefont {Taimre}}, \bibinfo {author} {\bibfnamefont {K.}~\bibnamefont {Bertling}}, \ and\ \bibinfo {author} {\bibfnamefont {A.~D.}\ \bibnamefont {Rakić}},\ }\bibfield  {title} {\enquote {\bibinfo {title} {Frequency combs in quantum cascade lasers: An overview of modeling and experiments},}\ }\href {\doibase https://doi.org/10.1063/5.0134539} {\bibfield  {journal} {\bibinfo  {journal} {APL Photonics}\ }\textbf {\bibinfo {volume} {8}},\ \bibinfo {pages} {020902} (\bibinfo {year} {2023}{\natexlab{a}})}\BibitemShut {NoStop}%
\bibitem [{\citenamefont {Piccardo}\ and\ \citenamefont {Capasso}(2022)}]{PiccardoReview}%
  \BibitemOpen
  \bibfield  {author} {\bibinfo {author} {\bibfnamefont {M.}~\bibnamefont {Piccardo}}\ and\ \bibinfo {author} {\bibfnamefont {F.}~\bibnamefont {Capasso}},\ }\bibfield  {title} {\enquote {\bibinfo {title} {Laser frequency combs with fast gain recovery: Physics and applications},}\ }\href {\doibase https://doi.org/10.1002/lpor.202100403} {\bibfield  {journal} {\bibinfo  {journal} {Laser \& Photonics Reviews}\ }\textbf {\bibinfo {volume} {16}},\ \bibinfo {pages} {2100403} (\bibinfo {year} {2022})}\BibitemShut {NoStop}%
\bibitem [{\citenamefont {Faist}\ \emph {et~al.}(2016)\citenamefont {Faist}, \citenamefont {Villares}, \citenamefont {Scalari}, \citenamefont {Rösch}, \citenamefont {Bonzon}, \citenamefont {Hugi},\ and\ \citenamefont {Beck}}]{Faist_2016}%
  \BibitemOpen
  \bibfield  {author} {\bibinfo {author} {\bibfnamefont {J.}~\bibnamefont {Faist}}, \bibinfo {author} {\bibfnamefont {G.}~\bibnamefont {Villares}}, \bibinfo {author} {\bibfnamefont {G.}~\bibnamefont {Scalari}}, \bibinfo {author} {\bibfnamefont {M.}~\bibnamefont {Rösch}}, \bibinfo {author} {\bibfnamefont {C.}~\bibnamefont {Bonzon}}, \bibinfo {author} {\bibfnamefont {A.}~\bibnamefont {Hugi}}, \ and\ \bibinfo {author} {\bibfnamefont {M.}~\bibnamefont {Beck}},\ }\bibfield  {title} {\enquote {\bibinfo {title} {Quantum cascade laser frequency combs},}\ }\href {\doibase doi:10.1515/nanoph-2016-0015} {\bibfield  {journal} {\bibinfo  {journal} {Nanophotonics}\ }\textbf {\bibinfo {volume} {5}},\ \bibinfo {pages} {272--291} (\bibinfo {year} {2016})}\BibitemShut {NoStop}%
\bibitem [{\citenamefont {Villares}\ \emph {et~al.}(2014)\citenamefont {Villares}, \citenamefont {Hugi}, \citenamefont {Blaser},\ and\ \citenamefont {Faist}}]{Villares_2014}%
  \BibitemOpen
  \bibfield  {author} {\bibinfo {author} {\bibfnamefont {G.}~\bibnamefont {Villares}}, \bibinfo {author} {\bibfnamefont {A.}~\bibnamefont {Hugi}}, \bibinfo {author} {\bibfnamefont {S.}~\bibnamefont {Blaser}}, \ and\ \bibinfo {author} {\bibfnamefont {J.}~\bibnamefont {Faist}},\ }\bibfield  {title} {\enquote {\bibinfo {title} {Dual-comb spectroscopy based on quantum-cascade-laser frequency combs},}\ }\href {\doibase 10.1038/ncomms6192} {\bibfield  {journal} {\bibinfo  {journal} {Nature Communications}\ }\textbf {\bibinfo {volume} {5}},\ \bibinfo {pages} {5192} (\bibinfo {year} {2014})}\BibitemShut {NoStop}%
\bibitem [{\citenamefont {Corrias}\ \emph {et~al.}(2022)\citenamefont {Corrias}, \citenamefont {Gabbrielli}, \citenamefont {Natale}, \citenamefont {Consolino},\ and\ \citenamefont {Cappelli}}]{Corrias}%
  \BibitemOpen
  \bibfield  {author} {\bibinfo {author} {\bibfnamefont {N.}~\bibnamefont {Corrias}}, \bibinfo {author} {\bibfnamefont {T.}~\bibnamefont {Gabbrielli}}, \bibinfo {author} {\bibfnamefont {P.~D.}\ \bibnamefont {Natale}}, \bibinfo {author} {\bibfnamefont {L.}~\bibnamefont {Consolino}}, \ and\ \bibinfo {author} {\bibfnamefont {F.}~\bibnamefont {Cappelli}},\ }\bibfield  {title} {\enquote {\bibinfo {title} {Analog {FM} free-space optical communication based on a mid-infrared quantum cascade laser frequency comb},}\ }\href {\doibase 10.1364/OE.443483} {\bibfield  {journal} {\bibinfo  {journal} {Opt. Express}\ }\textbf {\bibinfo {volume} {30}},\ \bibinfo {pages} {10217--10228} (\bibinfo {year} {2022})}\BibitemShut {NoStop}%
\bibitem [{\citenamefont {Kazakov}\ \emph {et~al.}(2021)\citenamefont {Kazakov}, \citenamefont {Opa\v{c}ak}, \citenamefont {Beiser}, \citenamefont {Belyanin}, \citenamefont {Schwarz}, \citenamefont {Piccardo},\ and\ \citenamefont {Capasso}}]{Kazakov21}%
  \BibitemOpen
  \bibfield  {author} {\bibinfo {author} {\bibfnamefont {D.}~\bibnamefont {Kazakov}}, \bibinfo {author} {\bibfnamefont {N.}~\bibnamefont {Opa\v{c}ak}}, \bibinfo {author} {\bibfnamefont {M.}~\bibnamefont {Beiser}}, \bibinfo {author} {\bibfnamefont {A.}~\bibnamefont {Belyanin}}, \bibinfo {author} {\bibfnamefont {B.}~\bibnamefont {Schwarz}}, \bibinfo {author} {\bibfnamefont {M.}~\bibnamefont {Piccardo}}, \ and\ \bibinfo {author} {\bibfnamefont {F.}~\bibnamefont {Capasso}},\ }\bibfield  {title} {\enquote {\bibinfo {title} {Defect-engineered ring laser harmonic frequency combs},}\ }\href {\doibase 10.1364/OPTICA.430896} {\bibfield  {journal} {\bibinfo  {journal} {Optica}\ }\textbf {\bibinfo {volume} {8}},\ \bibinfo {pages} {1277--1280} (\bibinfo {year} {2021})}\BibitemShut {NoStop}%
\bibitem [{\citenamefont {Forrer}\ \emph {et~al.}(2021)\citenamefont {Forrer}, \citenamefont {Wang}, \citenamefont {Beck}, \citenamefont {Belyanin}, \citenamefont {Faist},\ and\ \citenamefont {Scalari}}]{ForrerHFC}%
  \BibitemOpen
  \bibfield  {author} {\bibinfo {author} {\bibfnamefont {A.}~\bibnamefont {Forrer}}, \bibinfo {author} {\bibfnamefont {Y.}~\bibnamefont {Wang}}, \bibinfo {author} {\bibfnamefont {M.}~\bibnamefont {Beck}}, \bibinfo {author} {\bibfnamefont {A.}~\bibnamefont {Belyanin}}, \bibinfo {author} {\bibfnamefont {J.}~\bibnamefont {Faist}}, \ and\ \bibinfo {author} {\bibfnamefont {G.}~\bibnamefont {Scalari}},\ }\bibfield  {title} {\enquote {\bibinfo {title} {Self-starting harmonic comb emission in {TH}z quantum cascade lasers},}\ }\href@noop {} {\bibfield  {journal} {\bibinfo  {journal} {Applied Physics Letters}\ }\textbf {\bibinfo {volume} {118}},\ \bibinfo {pages} {131112} (\bibinfo {year} {2021})}\BibitemShut {NoStop}%
\bibitem [{\citenamefont {Wang}\ \emph {et~al.}(2020)\citenamefont {Wang}, \citenamefont {Pistore}, \citenamefont {Riesch}, \citenamefont {Nong}, \citenamefont {Vigneron}, \citenamefont {Colombelli}, \citenamefont {Parillaud}, \citenamefont {Mangeney}, \citenamefont {Tignon}, \citenamefont {Jirauschek},\ and\ \citenamefont {Dhillon}}]{Dhillon1}%
  \BibitemOpen
  \bibfield  {author} {\bibinfo {author} {\bibfnamefont {F.}~\bibnamefont {Wang}}, \bibinfo {author} {\bibfnamefont {V.}~\bibnamefont {Pistore}}, \bibinfo {author} {\bibfnamefont {M.}~\bibnamefont {Riesch}}, \bibinfo {author} {\bibfnamefont {H.}~\bibnamefont {Nong}}, \bibinfo {author} {\bibfnamefont {P.-B.}\ \bibnamefont {Vigneron}}, \bibinfo {author} {\bibfnamefont {R.}~\bibnamefont {Colombelli}}, \bibinfo {author} {\bibfnamefont {O.}~\bibnamefont {Parillaud}}, \bibinfo {author} {\bibfnamefont {J.}~\bibnamefont {Mangeney}}, \bibinfo {author} {\bibfnamefont {J.}~\bibnamefont {Tignon}}, \bibinfo {author} {\bibfnamefont {C.}~\bibnamefont {Jirauschek}}, \ and\ \bibinfo {author} {\bibfnamefont {S.~S.}\ \bibnamefont {Dhillon}},\ }\bibfield  {title} {\enquote {\bibinfo {title} {Ultrafast response of harmonic modelocked {TH}z lasers},}\ }\href@noop {} {\bibfield  {journal} {\bibinfo  {journal} {Light: Science {\&} Applications}\ }\textbf {\bibinfo {volume} {9}},\ \bibinfo {pages} {51} (\bibinfo {year}
  {2020})}\BibitemShut {NoStop}%
\bibitem [{\citenamefont {Kazakov}\ \emph {et~al.}(2017)\citenamefont {Kazakov}, \citenamefont {Piccardo}, \citenamefont {Wang}, \citenamefont {Chevalier}, \citenamefont {Mansuripur}, \citenamefont {Xie}, \citenamefont {Zah}, \citenamefont {Lascola}, \citenamefont {Belyanin},\ and\ \citenamefont {Capasso}}]{Kazakov2017}%
  \BibitemOpen
  \bibfield  {author} {\bibinfo {author} {\bibfnamefont {D.}~\bibnamefont {Kazakov}}, \bibinfo {author} {\bibfnamefont {M.}~\bibnamefont {Piccardo}}, \bibinfo {author} {\bibfnamefont {Y.}~\bibnamefont {Wang}}, \bibinfo {author} {\bibfnamefont {P.}~\bibnamefont {Chevalier}}, \bibinfo {author} {\bibfnamefont {T.~S.}\ \bibnamefont {Mansuripur}}, \bibinfo {author} {\bibfnamefont {F.}~\bibnamefont {Xie}}, \bibinfo {author} {\bibfnamefont {C.-e.}\ \bibnamefont {Zah}}, \bibinfo {author} {\bibfnamefont {K.}~\bibnamefont {Lascola}}, \bibinfo {author} {\bibfnamefont {A.}~\bibnamefont {Belyanin}}, \ and\ \bibinfo {author} {\bibfnamefont {F.}~\bibnamefont {Capasso}},\ }\bibfield  {title} {\enquote {\bibinfo {title} {Self-starting harmonic frequency comb generation in a quantum cascade laser},}\ }\href@noop {} {\bibfield  {journal} {\bibinfo  {journal} {Nature Photonics}\ }\textbf {\bibinfo {volume} {11}},\ \bibinfo {pages} {789--792} (\bibinfo {year} {2017})}\BibitemShut {NoStop}%
\bibitem [{\citenamefont {Piccardo}\ \emph {et~al.}(2018{\natexlab{a}})\citenamefont {Piccardo}, \citenamefont {Chevalier}, \citenamefont {Mansuripur}, \citenamefont {Kazakov}, \citenamefont {Wang}, \citenamefont {Rubin}, \citenamefont {Meadowcroft}, \citenamefont {Belyanin},\ and\ \citenamefont {Capasso}}]{PiccardoHFCOptex}%
  \BibitemOpen
  \bibfield  {author} {\bibinfo {author} {\bibfnamefont {M.}~\bibnamefont {Piccardo}}, \bibinfo {author} {\bibfnamefont {P.}~\bibnamefont {Chevalier}}, \bibinfo {author} {\bibfnamefont {T.~S.}\ \bibnamefont {Mansuripur}}, \bibinfo {author} {\bibfnamefont {D.}~\bibnamefont {Kazakov}}, \bibinfo {author} {\bibfnamefont {Y.}~\bibnamefont {Wang}}, \bibinfo {author} {\bibfnamefont {N.~A.}\ \bibnamefont {Rubin}}, \bibinfo {author} {\bibfnamefont {L.}~\bibnamefont {Meadowcroft}}, \bibinfo {author} {\bibfnamefont {A.}~\bibnamefont {Belyanin}}, \ and\ \bibinfo {author} {\bibfnamefont {F.}~\bibnamefont {Capasso}},\ }\bibfield  {title} {\enquote {\bibinfo {title} {The harmonic state of quantum cascade lasers: origin, control, and prospective applications},}\ }\href@noop {} {\bibfield  {journal} {\bibinfo  {journal} {Opt. Express}\ }\textbf {\bibinfo {volume} {26}},\ \bibinfo {pages} {9464--9483} (\bibinfo {year} {2018}{\natexlab{a}})}\BibitemShut {NoStop}%
\bibitem [{\citenamefont {Silvestri}\ \emph {et~al.}(2023{\natexlab{b}})\citenamefont {Silvestri}, \citenamefont {Qi}, \citenamefont {Taimre},\ and\ \citenamefont {Raki\ifmmode~\acute{c}\else \'{c}\fi{}}}]{Silvestri23PRA}%
  \BibitemOpen
  \bibfield  {author} {\bibinfo {author} {\bibfnamefont {C.}~\bibnamefont {Silvestri}}, \bibinfo {author} {\bibfnamefont {X.}~\bibnamefont {Qi}}, \bibinfo {author} {\bibfnamefont {T.}~\bibnamefont {Taimre}}, \ and\ \bibinfo {author} {\bibfnamefont {A.~D.}\ \bibnamefont {Raki\ifmmode~\acute{c}\else \'{c}\fi{}}},\ }\bibfield  {title} {\enquote {\bibinfo {title} {Harmonic active mode locking in terahertz quantum cascade lasers},}\ }\href {\doibase 10.1103/PhysRevA.108.013501} {\bibfield  {journal} {\bibinfo  {journal} {Phys. Rev. A}\ }\textbf {\bibinfo {volume} {108}},\ \bibinfo {pages} {013501} (\bibinfo {year} {2023}{\natexlab{b}})}\BibitemShut {NoStop}%
\bibitem [{\citenamefont {Piccardo}\ \emph {et~al.}(2018{\natexlab{b}})\citenamefont {Piccardo}, \citenamefont {Chevalier}, \citenamefont {Anand}, \citenamefont {Wang}, \citenamefont {Kazakov}, \citenamefont {Mejia}, \citenamefont {Xie}, \citenamefont {Lascola}, \citenamefont {Belyanin},\ and\ \citenamefont {Capasso}}]{PiccardoOptical}%
  \BibitemOpen
  \bibfield  {author} {\bibinfo {author} {\bibfnamefont {M.}~\bibnamefont {Piccardo}}, \bibinfo {author} {\bibfnamefont {P.}~\bibnamefont {Chevalier}}, \bibinfo {author} {\bibfnamefont {S.}~\bibnamefont {Anand}}, \bibinfo {author} {\bibfnamefont {Y.}~\bibnamefont {Wang}}, \bibinfo {author} {\bibfnamefont {D.}~\bibnamefont {Kazakov}}, \bibinfo {author} {\bibfnamefont {E.~A.}\ \bibnamefont {Mejia}}, \bibinfo {author} {\bibfnamefont {F.}~\bibnamefont {Xie}}, \bibinfo {author} {\bibfnamefont {K.}~\bibnamefont {Lascola}}, \bibinfo {author} {\bibfnamefont {A.}~\bibnamefont {Belyanin}}, \ and\ \bibinfo {author} {\bibfnamefont {F.}~\bibnamefont {Capasso}},\ }\bibfield  {title} {\enquote {\bibinfo {title} {Widely tunable harmonic frequency comb in a quantum cascade laser},}\ }\href@noop {} {\bibfield  {journal} {\bibinfo  {journal} {Applied Physics Letters}\ }\textbf {\bibinfo {volume} {113}},\ \bibinfo {pages} {031104} (\bibinfo {year} {2018}{\natexlab{b}})}\BibitemShut {NoStop}%
\bibitem [{\citenamefont {Riccardi}\ \emph {et~al.}(2024)\citenamefont {Riccardi}, \citenamefont {Guerrero}, \citenamefont {Pistore}, \citenamefont {Seitner}, \citenamefont {Jirauschek}, \citenamefont {Li}, \citenamefont {Davies}, \citenamefont {Linfield},\ and\ \citenamefont {Vitiello}}]{VitielloHFC1}%
  \BibitemOpen
  \bibfield  {author} {\bibinfo {author} {\bibfnamefont {E.}~\bibnamefont {Riccardi}}, \bibinfo {author} {\bibfnamefont {M.~A.~J.}\ \bibnamefont {Guerrero}}, \bibinfo {author} {\bibfnamefont {V.}~\bibnamefont {Pistore}}, \bibinfo {author} {\bibfnamefont {L.}~\bibnamefont {Seitner}}, \bibinfo {author} {\bibfnamefont {C.}~\bibnamefont {Jirauschek}}, \bibinfo {author} {\bibfnamefont {L.}~\bibnamefont {Li}}, \bibinfo {author} {\bibfnamefont {A.~G.}\ \bibnamefont {Davies}}, \bibinfo {author} {\bibfnamefont {E.~H.}\ \bibnamefont {Linfield}}, \ and\ \bibinfo {author} {\bibfnamefont {M.~S.}\ \bibnamefont {Vitiello}},\ }\bibfield  {title} {\enquote {\bibinfo {title} {Sculpting harmonic comb states in terahertz quantum cascade lasers by controlled engineering},}\ }\href {\doibase 10.1364/OPTICA.509929} {\bibfield  {journal} {\bibinfo  {journal} {Optica}\ }\textbf {\bibinfo {volume} {11}},\ \bibinfo {pages} {412--419} (\bibinfo {year} {2024})}\BibitemShut {NoStop}%
\bibitem [{\citenamefont {Guerrero}\ \emph {et~al.}(2024)\citenamefont {Guerrero}, \citenamefont {Arif}, \citenamefont {Sorba},\ and\ \citenamefont {Vitiello}}]{VitielloHFC2}%
  \BibitemOpen
  \bibfield  {author} {\bibinfo {author} {\bibfnamefont {M.~A.~J.}\ \bibnamefont {Guerrero}}, \bibinfo {author} {\bibfnamefont {O.}~\bibnamefont {Arif}}, \bibinfo {author} {\bibfnamefont {L.}~\bibnamefont {Sorba}}, \ and\ \bibinfo {author} {\bibfnamefont {M.~S.}\ \bibnamefont {Vitiello}},\ }\bibfield  {title} {\enquote {\bibinfo {title} {Harmonic quantum cascade laser terahertz frequency combs enabled by multilayer graphene top-cavity scatters},}\ }\href {\doibase doi:10.1515/nanoph-2023-0912} {\bibfield  {journal} {\bibinfo  {journal} {Nanophotonics}\ }\textbf {\bibinfo {volume} {13}},\ \bibinfo {pages} {1835--1841} (\bibinfo {year} {2024})}\BibitemShut {NoStop}%
\bibitem [{\citenamefont {Silvestri}\ \emph {et~al.}(2023{\natexlab{c}})\citenamefont {Silvestri}, \citenamefont {Qi}, \citenamefont {Taimre},\ and\ \citenamefont {Rakić}}]{Silvestrifb}%
  \BibitemOpen
  \bibfield  {author} {\bibinfo {author} {\bibfnamefont {C.}~\bibnamefont {Silvestri}}, \bibinfo {author} {\bibfnamefont {X.}~\bibnamefont {Qi}}, \bibinfo {author} {\bibfnamefont {T.}~\bibnamefont {Taimre}}, \ and\ \bibinfo {author} {\bibfnamefont {A.~D.}\ \bibnamefont {Rakić}},\ }\bibfield  {title} {\enquote {\bibinfo {title} {{Frequency combs induced by optical feedback and harmonic order tunability in quantum cascade lasers}},}\ }\href@noop {} {\bibfield  {journal} {\bibinfo  {journal} {APL Photonics}\ }\textbf {\bibinfo {volume} {8}},\ \bibinfo {pages} {116102} (\bibinfo {year} {2023}{\natexlab{c}})}\BibitemShut {NoStop}%
\bibitem [{\citenamefont {Paul}, \citenamefont {Beigang},\ and\ \citenamefont {Rahm}(2009)}]{Paul09}%
  \BibitemOpen
  \bibfield  {author} {\bibinfo {author} {\bibfnamefont {O.}~\bibnamefont {Paul}}, \bibinfo {author} {\bibfnamefont {R.}~\bibnamefont {Beigang}}, \ and\ \bibinfo {author} {\bibfnamefont {M.}~\bibnamefont {Rahm}},\ }\bibfield  {title} {\enquote {\bibinfo {title} {Highly selective terahertz bandpass filters based on trapped mode excitation},}\ }\href {\doibase 10.1364/OE.17.018590} {\bibfield  {journal} {\bibinfo  {journal} {Opt. Express}\ }\textbf {\bibinfo {volume} {17}},\ \bibinfo {pages} {18590--18595} (\bibinfo {year} {2009})}\BibitemShut {NoStop}%
\bibitem [{\citenamefont {Columbo}\ \emph {et~al.}(2018)\citenamefont {Columbo}, \citenamefont {Barbieri}, \citenamefont {Sirtori},\ and\ \citenamefont {Brambilla}}]{Columbo2018}%
  \BibitemOpen
  \bibfield  {author} {\bibinfo {author} {\bibfnamefont {L.~L.}\ \bibnamefont {Columbo}}, \bibinfo {author} {\bibfnamefont {S.}~\bibnamefont {Barbieri}}, \bibinfo {author} {\bibfnamefont {C.}~\bibnamefont {Sirtori}}, \ and\ \bibinfo {author} {\bibfnamefont {M.}~\bibnamefont {Brambilla}},\ }\bibfield  {title} {\enquote {\bibinfo {title} {Dynamics of a broad-band quantum cascade laser: from chaos to coherent dynamics and mode-locking},}\ }\href {http://opg.optica.org/oe/abstract.cfm?URI=oe-26-3-2829} {\bibfield  {journal} {\bibinfo  {journal} {Opt. Express}\ }\textbf {\bibinfo {volume} {26}},\ \bibinfo {pages} {2829--2847} (\bibinfo {year} {2018})}\BibitemShut {NoStop}%
\bibitem [{\citenamefont {Prati}\ and\ \citenamefont {Columbo}(2007)}]{columbo2007}%
  \BibitemOpen
  \bibfield  {author} {\bibinfo {author} {\bibfnamefont {F.}~\bibnamefont {Prati}}\ and\ \bibinfo {author} {\bibfnamefont {L.}~\bibnamefont {Columbo}},\ }\bibfield  {title} {\enquote {\bibinfo {title} {Long-wavelength instability in broad-area semiconductor lasers},}\ }\href@noop {} {\bibfield  {journal} {\bibinfo  {journal} {Phys. Rev. A}\ }\textbf {\bibinfo {volume} {75}},\ \bibinfo {pages} {053811} (\bibinfo {year} {2007})}\BibitemShut {NoStop}%
\bibitem [{\citenamefont {Silvestri}\ \emph {et~al.}(2022)\citenamefont {Silvestri}, \citenamefont {Qi}, \citenamefont {Taimre},\ and\ \citenamefont {Raki\ifmmode~\acute{c}\else \'{c}\fi{}}}]{Silvestri22}%
  \BibitemOpen
  \bibfield  {author} {\bibinfo {author} {\bibfnamefont {C.}~\bibnamefont {Silvestri}}, \bibinfo {author} {\bibfnamefont {X.}~\bibnamefont {Qi}}, \bibinfo {author} {\bibfnamefont {T.}~\bibnamefont {Taimre}}, \ and\ \bibinfo {author} {\bibfnamefont {A.~D.}\ \bibnamefont {Raki\ifmmode~\acute{c}\else \'{c}\fi{}}},\ }\bibfield  {title} {\enquote {\bibinfo {title} {Multimode dynamics of terahertz quantum cascade lasers: Spontaneous and actively induced generation of dense and harmonic coherent regimes},}\ }\href {\doibase 10.1103/PhysRevA.106.053526} {\bibfield  {journal} {\bibinfo  {journal} {Phys. Rev. A}\ }\textbf {\bibinfo {volume} {106}},\ \bibinfo {pages} {053526} (\bibinfo {year} {2022})}\BibitemShut {NoStop}%
\bibitem [{\citenamefont {Silvestri}(2022)}]{SilvestriThesis}%
  \BibitemOpen
  \bibfield  {author} {\bibinfo {author} {\bibfnamefont {C.}~\bibnamefont {Silvestri}},\ }\emph {\bibinfo {title} {Theory and modelization of Quantum Cascade Laser dynamics: comb formation, field structures and feedback-based imaging}},\ \href@noop {} {Ph.D. thesis},\ \bibinfo  {school} {Politecnico di Torino, Torino, Italy} (\bibinfo {year} {2022})\BibitemShut {NoStop}%
\bibitem [{\citenamefont {Silvestri}\ \emph {et~al.}(2020)\citenamefont {Silvestri}, \citenamefont {Columbo}, \citenamefont {Brambilla},\ and\ \citenamefont {Gioannini}}]{Silvestri20}%
  \BibitemOpen
  \bibfield  {author} {\bibinfo {author} {\bibfnamefont {C.}~\bibnamefont {Silvestri}}, \bibinfo {author} {\bibfnamefont {L.~L.}\ \bibnamefont {Columbo}}, \bibinfo {author} {\bibfnamefont {M.}~\bibnamefont {Brambilla}}, \ and\ \bibinfo {author} {\bibfnamefont {M.}~\bibnamefont {Gioannini}},\ }\bibfield  {title} {\enquote {\bibinfo {title} {Coherent multi-mode dynamics in a quantum cascade laser: amplitude- and frequency-modulated optical frequency combs},}\ }\href {\doibase 10.1364/OE.396481} {\bibfield  {journal} {\bibinfo  {journal} {Opt. Express}\ }\textbf {\bibinfo {volume} {28}},\ \bibinfo {pages} {23846--23861} (\bibinfo {year} {2020})}\BibitemShut {NoStop}%
\bibitem [{\citenamefont {{Digiorgio, Valerio}}\ \emph {et~al.}(2023)\citenamefont {{Digiorgio, Valerio}}, \citenamefont {{Senica, Urban}}, \citenamefont {{Micheletti, Paolo}}, \citenamefont {{Beck, Mattias}}, \citenamefont {{Faist, Jérôme}},\ and\ \citenamefont {{Scalari, Giacomo}}}]{comb_stability}%
  \BibitemOpen
  \bibfield  {author} {\bibinfo {author} {\bibnamefont {{Digiorgio, Valerio}}}, \bibinfo {author} {\bibnamefont {{Senica, Urban}}}, \bibinfo {author} {\bibnamefont {{Micheletti, Paolo}}}, \bibinfo {author} {\bibnamefont {{Beck, Mattias}}}, \bibinfo {author} {\bibnamefont {{Faist, Jérôme}}}, \ and\ \bibinfo {author} {\bibnamefont {{Scalari, Giacomo}}},\ }\bibfield  {title} {\enquote {\bibinfo {title} {Surface-emitting thz quantum cascade laser frequency comb with tunable external mirror dispersion compensation},}\ }\href {\doibase 10.1051/epjconf/202328707028} {\bibfield  {journal} {\bibinfo  {journal} {EPJ Web Conf.}\ }\textbf {\bibinfo {volume} {287}},\ \bibinfo {pages} {07028} (\bibinfo {year} {2023})}\BibitemShut {NoStop}%
\bibitem [{\citenamefont {Hillbrand}\ \emph {et~al.}(2018)\citenamefont {Hillbrand}, \citenamefont {Jouy}, \citenamefont {Beck},\ and\ \citenamefont {Faist}}]{Hillbrand18}%
  \BibitemOpen
  \bibfield  {author} {\bibinfo {author} {\bibfnamefont {J.}~\bibnamefont {Hillbrand}}, \bibinfo {author} {\bibfnamefont {P.}~\bibnamefont {Jouy}}, \bibinfo {author} {\bibfnamefont {M.}~\bibnamefont {Beck}}, \ and\ \bibinfo {author} {\bibfnamefont {J.}~\bibnamefont {Faist}},\ }\bibfield  {title} {\enquote {\bibinfo {title} {Tunable dispersion compensation of quantum cascade laser frequency combs},}\ }\href {\doibase 10.1364/OL.43.001746} {\bibfield  {journal} {\bibinfo  {journal} {Opt. Lett.}\ }\textbf {\bibinfo {volume} {43}},\ \bibinfo {pages} {1746--1749} (\bibinfo {year} {2018})}\BibitemShut {NoStop}%
\end{thebibliography}%

\end{document}